\documentclass{ws-jcsc}
\usepackage{graphicx}
\usepackage{multirow}
\usepackage{amsmath,amssymb,amsfonts}
\usepackage{mathrsfs}
\usepackage[title]{appendix}
\usepackage{xcolor}
\usepackage{textcomp}
\usepackage{manyfoot}
\usepackage{booktabs}
\usepackage{algorithm}

\usepackage{algorithmicx}
\usepackage{algpseudocode}
\usepackage{listings}

\usepackage{float}
\usepackage{array}
\usepackage{siunitx}
\usepackage{soul}

\usepackage{xcolor}

\usepackage{url,overcite}   

\newcommand{\purple}[1]{\textcolor{black}{#1}}

\begin{document}

\markboth{Ibrahim, Shokry, Siddhu, Bauer, Nassar, Henkel}{RISC-V ISE for MLC NVM}

\title{\purple{Retention-Aware RISC-V ISA Extension and Memory Controller on FPGA for MLC NVM}}

\author{Mina Ibrahim$^{\times}$\footnote{Both authors contributed equally}, Martel Shokry$^{\times*}$, Lokesh Siddhu$^{\S}$, Lars Bauer$^{\dag}$, Hassan Nassar$^{\S}$, Jörg Henkel$^{\S}$}

\address{$^{\times}$German University in Cairo (GUC), Egypt\\
minaibrahim2411@gmail.com, martelmegalaa46@gmail.com}

\address{$^{\S}$Karlsruhe Institute of Technology (KIT), Chair for Embedded Systems (CES)\\ Karlsruhe, Germany\\
lokesh.siddhu@kit.edu, hassan.nassar@kit.edu, henkel@kit.edu}

\address{$^{\dag}$Ubitium GmbH, Karlsruhe, Germany\\
lars.bauer42@gmx.de}

\markboth{FPGA-Based RISC-V and NVM Controller}{FPGA-Based RISC-V and NVM Controller}

\maketitle

\abstract{
Non-volatile memory (NVM) technologies, particularly Multi-Level Cell (MLC) NVMs, offer significant potential for increasing memory density. MLC NVMs provide a tradeoff between write latency and retention time, where faster writes/stores result in lower retention and slower writes yield higher retention. However, limited work has been done to validate and prototype NVM-based systems in hardware, leveraging this tradeoff at the system level.

In this paper, we present a novel memory controller architecture and a RISC-V instruction set extension to optimize MLC NVM write operations by balancing speed and retention time. Our custom NVM controller, built around a finite state machine with an AXI memory-mapped interface, efficiently manages read/write operations with enhanced burst transfers, minimizing latency. Additionally, we introduce a fast-store instruction in RISC-V to increasing write performance while addressing retention limitations. \purple{Further, we design a dedicated AXI slave peripheral that supports bit-significance-aware writes: critical bits (e.g., MSBs) are written using slower, high-retention writes, while non-critical bits (e.g., LSBs) use faster, low-retention writes to help enhance performance without compromising data reliability.} These enhancements are implemented in hardware on an FPGA platform. Experimental results show that our controller reduces hardware overhead by 30\% compared to conventional designs, and the fast-store instruction improves performance by over 7\% for streaming workloads with less than 0.08\% hardware overhead. \purple{The bit-wise AXI peripheral has a LUT utilization staying below 3.5\% even for 64×64 matrices, and under 1\% for 32×32 sizes, making it viable for integration into larger SoCs.}
}

\section{Introduction}\label{sec:intro}
Advancements in manufacturing technologies have driven towards adopting non-volatile memories (NVM) that utilize resistive or magnetic properties, rather than charge-based DRAM, as main memories for embedded systems~\cite{pcm_review_2010}. NVMs offer key advantages, such as high density, low power consumption, and non-volatility, making them strong candidates to replace DRAM. Many NVM types can be enhanced using Multi-Level Cell (MLC) technology~\cite{sttramMLC1,anv-puf}, which allows multiple bits to be stored in a single cell. For instance, in Phase Change Memory (PCM), bits are stored at different resistance levels. However, precise resistance (or any other physical property) control in MLC requires iterative write algorithms, which degrade write performance. One suggested approach to speed up the write process involves reducing the number of SET iterations, but this shortens data retention, leading to refreshing to maintain data integrity~\cite{zhang2019quick,siddhu2023swift, priya2020enhancing, mittal2016reliability}. Increased refreshes reduces memory lifespan and increases energy consumption~\cite{qureshi2009enhancing,nvmWear23,stt_mram_endurance_study,stt_mram_endurance_tsmc}.

Moreover, as policies for NVMs become more complex, prototyping and validating these techniques on real hardware is essential for achieving reliable, timely results—unlike simulation-based methods that may not fully capture hardware details and require long simulation runs. For hardware interfacing with NVMs, in terms of the processor selection, RISC-V~\cite{riscv3,ibrahim2024fpga} is an ideal choice due to its open-source nature and flexibility. Unlike proprietary instruction set architectures (ISAs), RISC-V allows developers to customize and extend the architecture to meet specific needs, enabling innovation in hardware design~\cite{riscv1}. This open-source approach not only reduces licensing costs but also creates a collaborative ecosystem where academia, industry, and individual developers can contribute to and benefit from a shared pool of resources. Consequently, RISC-V is gaining popularity in embedded systems, IoT, and specialized processors, where adaptability and cost-efficiency are critical~\cite{riscv2}.

\begin{figure}
    \centering
    \includegraphics[width=0.65\linewidth]{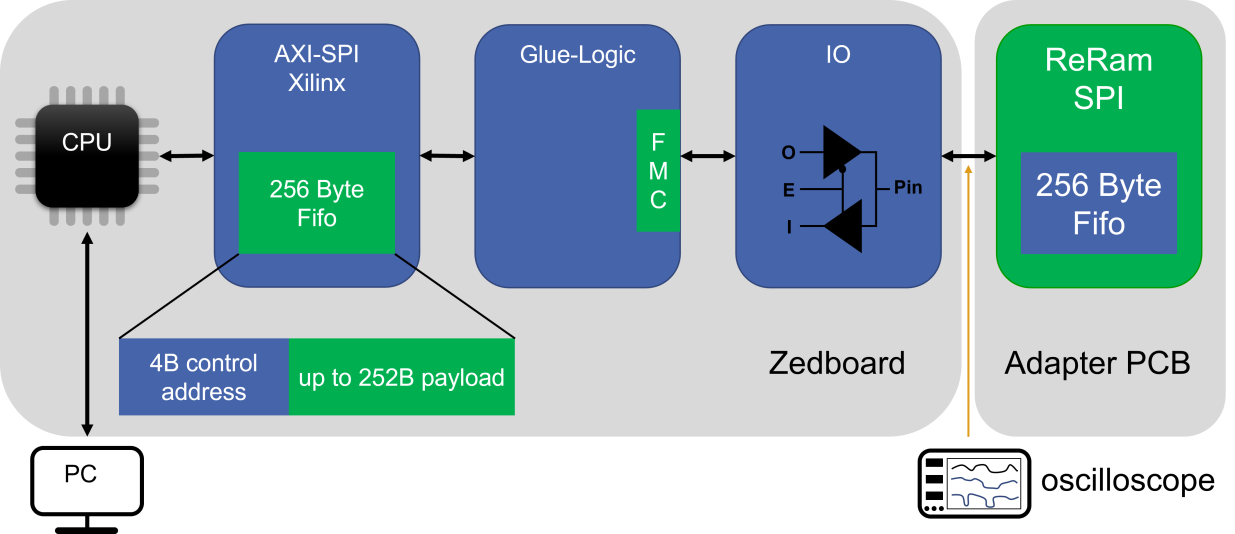}
    \caption{NVM Memory controller and extension}
    \label{fig:setup}
\end{figure}

FPGAs are an ideal candidate to implement RISC-V~\cite{riscv1}.
As RISC-V supports instruction set extensions, FPGAs can be used to implement hardware accelerators for the new instructions~\cite{nassar_icm_23, benhadjyoussef2015enhancing}.
In this paper, we target implementing a RISC-V on FPGA customized for NVM.
We aim to implement an extension instruction for RISC-V to handle an extra write mode of MLC for NVM.
Moreover, as Figure~\ref{fig:setup} shows, we target to build our own NVM-specific memory controller on FPGA to be able to interface several commercial NVM chips. \purple{We extend the memory controller to selectively apply different write modes: critical data bits (e.g., MSBs) are written using slow, high-retention operations, while less critical bits (e.g., LSBs) use fast, low-retention writes—improving overall performance while preserving data reliability.}

In summary our contributions are as follows:
\begin{itemize}
    \item We create a memory controller to connect NVMs with an FPGA via a custom FMC adapter. Using an FSM and an AXI memory-mapped interface, it efficiently manages read/write operations with burst transfers, allowing precise NVM performance.
    \item We propose utilizing the trade-off between write latency and retention time in NVMs by extending the RISC-V instruction set to include a new fast-store instruction.
    \item \purple{We propose a retention-aware write architecture that distinguishes bit significance: critical bits (e.g., MSBs) are written with longer retention, while non-critical bits (e.g., LSBs) use faster, short-retention writes—managed via a dedicated AXI slave peripheral.}
\end{itemize}

\purple{The paper is organized as follows. Sections~\ref{sec:Background} and~\ref{sec:Related Work} provide the background and review related work. Section~\ref{sec:Methodology} details the proposed NVM memory controller, the architectural modifications to support multiple write modes, and the extension for bit-significance-aware operations. Section~\ref{sec:Results} presents the timing analysis and hardware overhead evaluation. Finally, Section~\ref{sec:Conclusion} concludes the paper.}

\section{Background}
\label{sec:Background}

\begin{figure}
    \centering
    \includegraphics[width=0.65\linewidth]{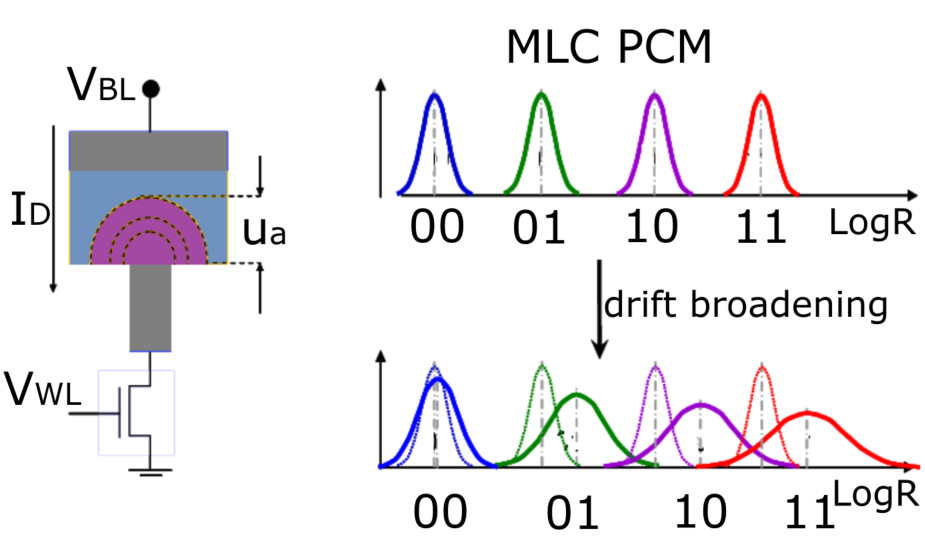}
    \caption{Drift resistance problem}
    \label{fig:Drift resistance problem}
\end{figure}

Non-volatile memory (NVM) technology, has evolved to offer increased storage capacity by enabling the storage of multiple digital bits within a single cell, known as Multi-Level Cell (MLC). However, MLC faces challenges related to retention time due to resistance drift, which can lead to errors in reading data and necessitate periodic refresh operations. Figure~\ref{fig:Drift resistance problem}, shows as an example the resistance drift in Phase Change Memory (PCM)~\cite{pcm_review_2010} which is a key issue for NVM. Research highlights a trade-off between write latency and retention time (Table~\ref{tab:Trade-off between Write Latency and Retention Time of MLC PCM}): write operations with more SET iterations and lower currents extend retention time but increase write latency, while fewer SET iterations improve write performance but reduce retention time, leading to more frequent refreshes. This trade-off also negatively affects the longevity of MLC NVM~\cite{stt_mram_endurance_study}. MLC offers greater storage capacity compared to Single-Level Cell (SLC), but it has a notable limitation: a shorter retention time, often lasting only a few hours due to resistance drift. As a result, MLC requires more frequent refresh operations than DRAM. Additionally, the precision of write operations directly affects retention time. Higher precision writes, involving more SET iterations, extend retention time but increase write latency. This creates a trade-off between write latency and retention in MLC. Continuing with the PCM example, a write operation with 7 SET iterations achieves a retention time of 3054.9 seconds, while one with 3 SET iterations results in a much shorter retention time of just 2.01 seconds as shown in Table~\ref{tab:Trade-off between Write Latency and Retention Time of MLC PCM}.
In addition to PCM, other NVM technologies exist such as ReRAM~\cite{rram_phys_model}, FRAM, MRAM~\cite{mram_fram}, and STTRAM~\cite{sttramMLC1}.

\purple{Due to resistance drift in MLC PCM, there is a trade-off between write latency and retention time. Further, in many applications, the most significant bits (MSBs) are more critical than the least significant bits (LSBs). Slower, high-retention writes can extend retention time but increase latency, while faster, low-retention writes improve performance but shorten retention. This trade-off complicates MLC-based NVM design. A potential solution is to apply slower writes to MSBs for better retention and faster writes to LSBs for improved performance, balancing both reliability and efficiency.}

\begin{table}
    \centering
    \caption{Trade-off between Write Latency and Retention Time of MLC PCM, based on~\cite{zhang2019quick}.}
    \begin{tabular}{|c|c|c|c|c|}
      \hline
       \textbf{Write Type} & \textbf{Latency} & \textbf{Retention} & \textbf{Current} & \textbf{Norm. Energy} \\
       
        \hline
        3-SETs-Write & 550\,ns & 2.01\,s & 42\,\(\mu\)A & 0.84\\
        \hline
        4-SETs-Write & 700\,ns & 24.05\,s & 37\,\(\mu\)A & 0.869\\
        \hline
        5-SETs-Write & 850\,ns & 104.4\,s & 35\,\(\mu\)A & 0.972\\
        \hline
        6-SETs-Write & 1000\,ns & 991.4\,s & 32\,\(\mu\)A & 0.975\\
        \hline
        7-SETs-Write & 1150\,ns & 3054.9\,s & 30\,\(\mu\)A & 1.00\\
        \hline
    \end{tabular}
    \label{tab:Trade-off between Write Latency and Retention Time of MLC PCM}
\end{table}

\section{Related Work}
\label{sec:Related Work}

\purple{Many efforts have addressed performance in Multi-Level Cell (MLC) Phase Change Memory (PCM) by utilizing trade-offs between write latency and retention time. The Quick-and-Dirty (QnD) architecture~\cite{zhang2019quick} issues fast, low-retention writes during high-demand periods and refreshes data in idle times, boosting system performance while maintaining reasonable memory lifespan with minimal changes to the memory controller. Li et al.~\cite{li2013compiler} present a compiler-directed dual-write scheme that optimizes the selection of write modes for improved performance. Pan et al.~\cite{pan20143m} explore task scheduling strategies that leverage fast and slow write modes, achieving notable energy savings in embedded systems. This line of work is further extended by Pan et al.~\cite{pan2017exploiting}, who develop a retention-time prediction model to dynamically select the optimal write mode for each memory access.}

\purple{Qiu et al.~\cite{qiu2015write} propose a write mode-aware loop tiling approach, which maximizes the use of fast writes by reducing retention requirements, thus enhancing both performance and memory endurance. Kim et al.~\cite{kim2016improving} improve write performance by controlling the target resistance distributions during programming, enabling faster and more reliable MLC PCM writes.
}

\purple{Retention-aware optimization has also been explored. Khwa et al.~\cite{khwa2016retention} introduce a metric for evaluating programming strategies in MLC PCM, allowing designers to quantitatively analyze the trade-offs between write speed and data retention. Swift-CNN~\cite{siddhu2023swift} adapts PCM write modes to the retention needs of different layers in convolutional neural networks, reducing both inference and training times while lowering energy consumption.}

\purple{While these works have made significant progress, most rely on simulation for evaluation, which can be slow and susceptible to modeling inaccuracies. In contrast, we develop a hardware/FPGA-based platform for reliable and faster evaluation and emulation, enabling more accurate characterization of MLC PCM performance and reliability trade-offs.
}

\section{Extending RISC-V and Implementing our Memory Controller}
\label{sec:Methodology}

Using NVM effectively necessitates the development of an NVM memory controller for a RISC-V platform, including the extension of the instruction set for MLC operations. The memory controller involves a custom FSM for NVM interfacing. While the new store instructions improve memory operations, they required changes to hardware (memory controller and cache) and software tools (compiler and assembler).

    \begin{figure}[t]
        \centering
        \includegraphics[width=0.65\linewidth]{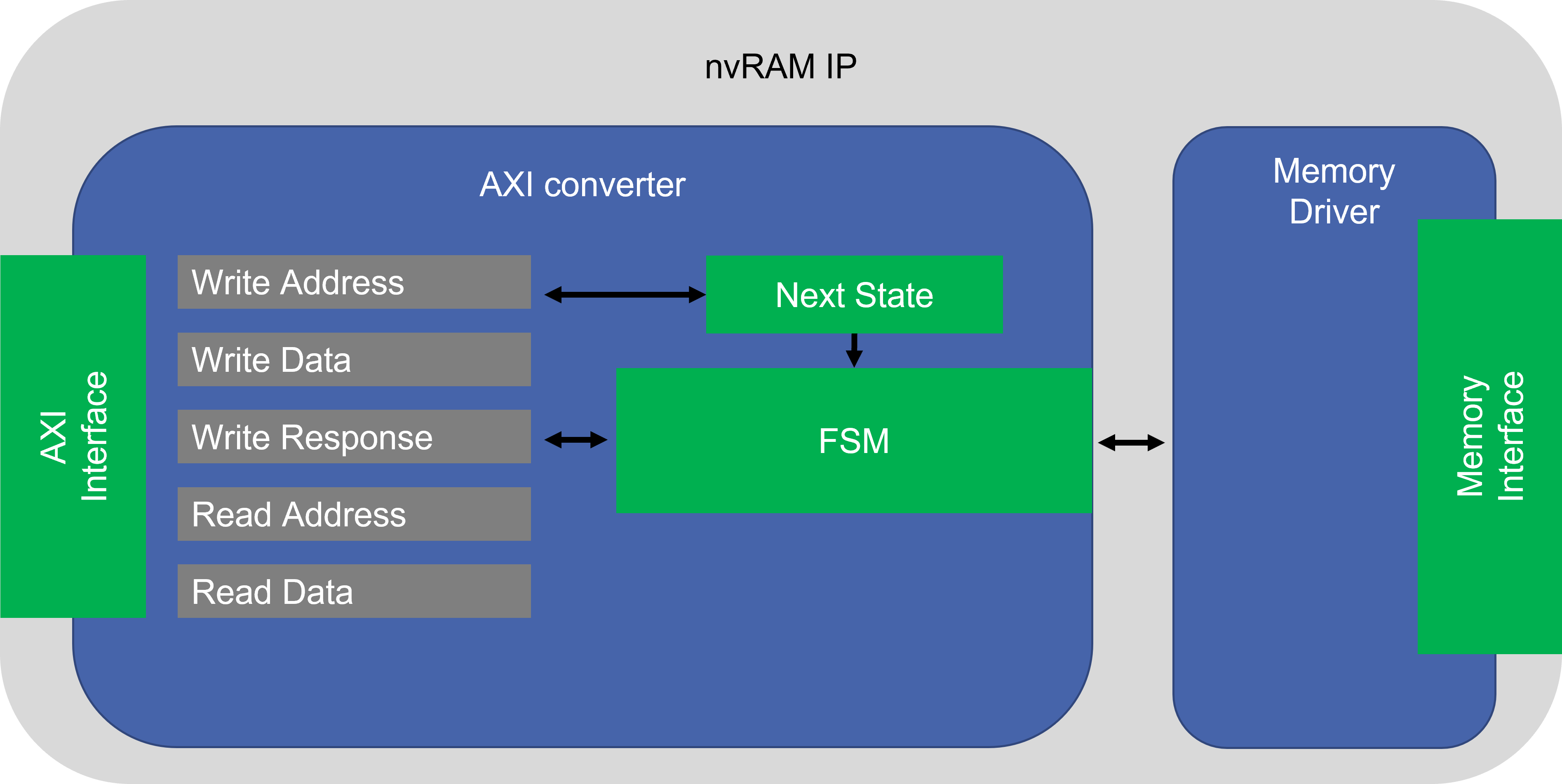}
        \caption{Design of memory controller with AXI interface}
        \label{fig: Design of memory controller with AXI interface}
    \end{figure}

\subsection{Memory Controller for NVM} 
\label{sub_sec:MC for NVM}
The memory controller designed in this work is designed to be able to accomodate different commercial NVM chips from different suppliers. 
Our NVM memory controller consists mainly of two parts as Figure~\ref{fig: Design of memory controller with AXI interface} shows.
First, a component to communicate with the processor via an AXI interface.
Second, a memory driver to interface with our developed PCB containing the commercial chips. 
The component communicating with the processor uses an FSM to regulate the communication between the processor and the different NVM chips.

\begin{figure}
  \centering
  \includegraphics[width=0.85\linewidth]{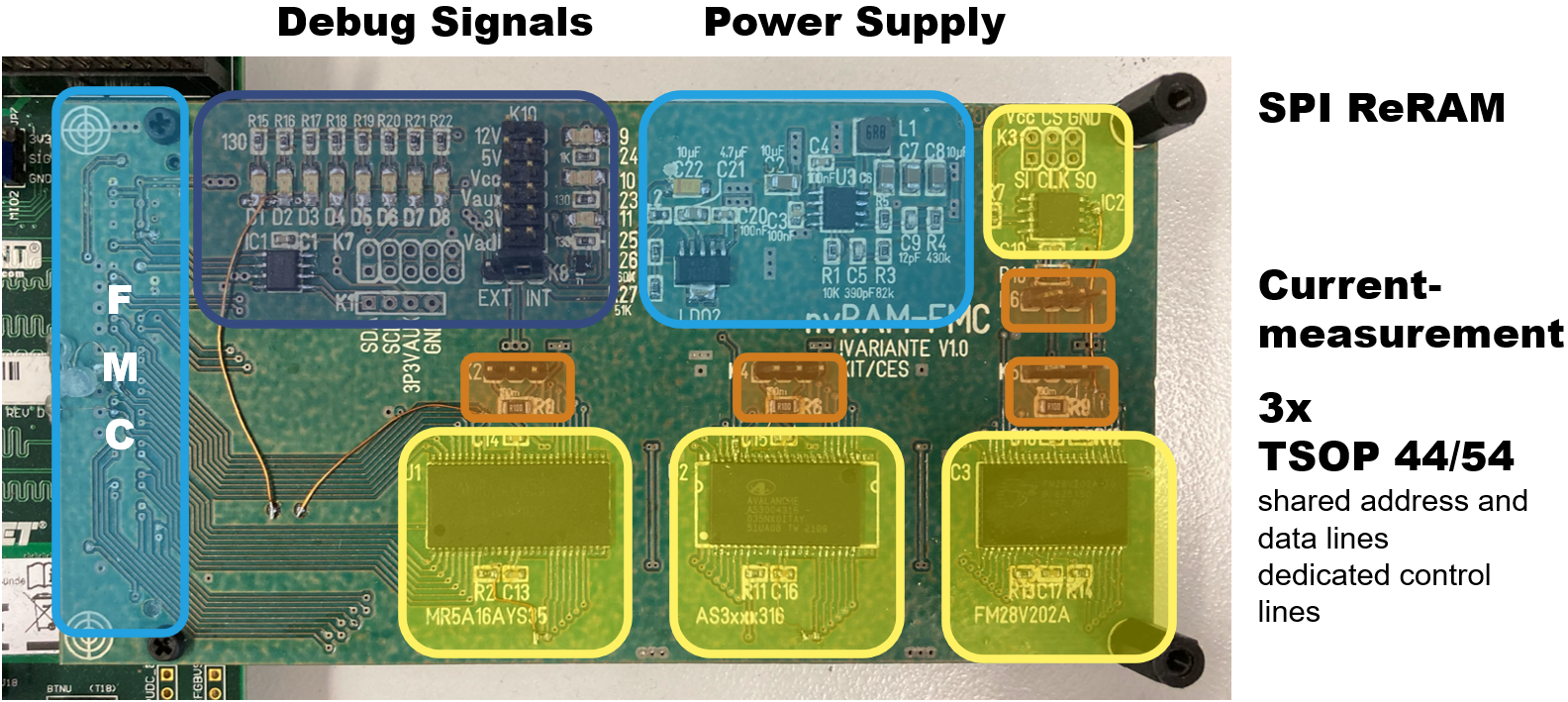}
  \caption[PCB 3.3V]{FMC Adapter board in the 3.3V version with color-coded sections}
  \label{fig:PCB3V}
\end{figure}

\begin{figure}
    \centering
    \includegraphics[width=0.65\linewidth]{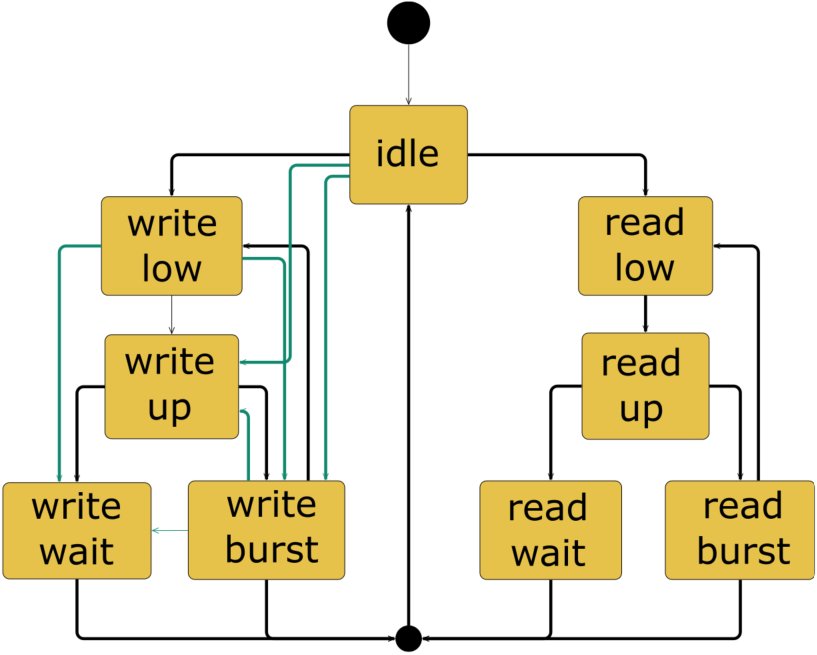}
    \caption{State Diagram For AXI FSM}
    \label{fig:State Diagram}
\end{figure}

The FSM is designed to control how the memory controller interacts with the non-volatile memory (NVM). It follows a structured approach where each memory access operation (read or write) is broken into smaller, timed phases. For instance:
\begin{itemize}
    \item Read Cycle: The FSM goes through the following phases—enable (assert memory chip enable), wait (wait for data to be valid), read (sample data), and end (finalize the operation).
    \item  Write Cycle: The FSM includes phases for enable, write (put data on the bus), and recovery (allow memory to stabilize).
\end{itemize}    
These phases ensure proper coordination between the FPGA and the NVM by accounting for delays and ensuring signals are stable before the next phase begins. The clock frequency is set at 100 MHz, so each FSM step corresponds to a 10 ns period.

The FSM also manages the AXI transactions. Since the non-volatile memories work at a 16-bit width and the AXI bus operates at a 32-bit width, the FSM splits the AXI data into two parts: the lower 16 bits and the upper 16 bits. Each part is processed sequentially in separate states (low, up) for both read and write operations.
State Diagram: The FSM starts in an idle state, then moves through states for reading and writing the lower 16 bits (read low, write low), and then for the upper 16 bits (read up, write up). The FSM returns to idle once the full 32-bit word is processed as shown in Figure~\ref{fig:State Diagram}.

The AXI interface was further optimized by introducing burst transfers, which allow multiple sequential memory accesses to be performed with a single address transmission. This reduces overhead and speeds up large data transfers, such as those involved in direct memory access (DMA) operations.

The FSM for AXI is designed to handle both the lower and upper 16 bits of a memory word sequentially. However, additional logic is introduced to skip unnecessary FSM states when, for example, only one part (either lower or upper) needs to be processed. This ensures that operations involving only partial data do not waste cycles. 
The AXI bus protocol automatically manages the synchronization between the CPU and the memory controller, removing the need for manual synchronization as in the earlier GPIO-based approach. This simplifies the software interaction with the memory controller and eliminates potential data corruption or mismatches which can be shown using the block diagram in Figure~\ref{fig: Design of memory controller with AXI interface}.

\subsubsection{PCB Layout} 
\label{sec:pcb_layout}

The memory controller interfaces with the NVM memories laid out on a PCB board. The PCB provides connections for current measurements are highlighted in orange in Figure~\ref{fig:PCB3V}, which shows the completed board. Current sense resistors are located near them. Yellow highlights mark the footprints and mounted nvRAMs, with the TSOP memory chips placed on the bottom half of the board. Although address and data lines are numbered differently, they share the same pin locations. Larger memory chips with more address lines are arranged to accommodate smaller chips on the same footprint. The connections follow a bus topology horizontally, routed vertically via vias. Signals are grouped into 8-bit blocks to facilitate future integration of level shifters between the FMC connector (blue, bottom) and the first memory chip.

All three nvRAMs share control signals: \textit{ChipSelect} (CS), \textit{OutputEnable} (OE), and \textit{WriteEnable} (WE), with identical supply pin locations. The differences lie in the pull-up resistor configurations during operation and startup. FRAM has an additional \textit{Sleep} pin, absent in newer MRAM revisions, where it remains unconnected.

The power supply section (light blue) is positioned opposite the nvRAMs to isolate high-frequency signals from the switching regulator. Between the power supply and FMC connector are debug LEDs and pin headers for spare signals. Control LEDs for \SI{12}{\volt}, \SI{5}{\volt}, and \SI{3.3}{\volt} supplies are also placed here. Generated and FMC-provided voltages are accessible via pin headers for debugging. A jumper (K8 EXIT-INT) allows switching between board-generated and external \SI{3.3}{\volt} power from the FMC connector, enabling external power use in case of failure or minimal assembly.

An I2C EEPROM near the FMC connector supports potential IPMI integration, storing board voltages or a unique identifier. It can also help distinguish between board versions during setup, such as identifying defects like the "FMC card is inserted" signal issue.

\subsection{RISC-V Extension for NVM Write Mode}  
\label{sub_sec:New Instruction}
After implementing our memory controller we had to extend RISC-V to support MLC with a new instruction.
In order to integrate the new instruction into the compiler toolchain, available opcodes were identified from the RISC-V instruction set manual~\cite{Waterman_Asanovic_2017}. Four opcodes were selected to accommodate varying store granularities, ranging from byte to double-word operations. These opcodes were located under the \textit{S-type} main opcode, with minor opcodes ranging from four to seven, where opcode four corresponds to the fast store byte instruction, and opcode seven corresponds to the fast store double-word instruction.

\begin{figure}
    \centering
    \includegraphics[width=0.65\linewidth]{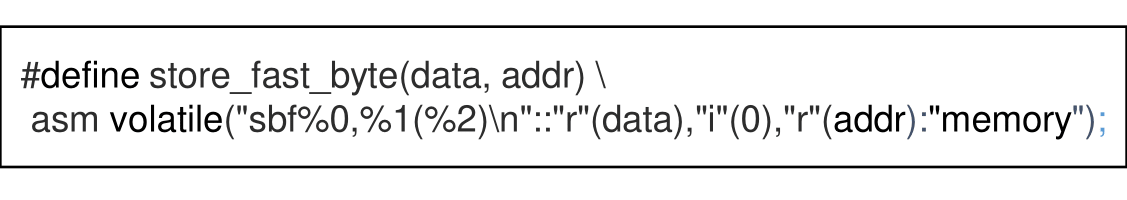}
    \caption{Instruction Format}
    \label{fig:format}
\end{figure}

Four new C macros were implemented to support all data granularities. Store byte fast instruction is shown in Figure~\ref{fig:format} to illustrate the used format. These macros utilize inline assembly to directly pass the instructions to the assembler. The function parameters consist of the data to be stored and the target address. The \textit{volatile} qualifier is employed to ensure that the assembler does not eliminate the instructions due to their lack of output. Additionally, the \textit{'memory'} clobber is used to prevent the compiler from reordering instructions around new instructions. The instruction format follows the S-type structure, similar to the standard store format, as it requires the same parameters. It needs a data register, a register containing the address, and another for the immediate offset, which is used to compute the final address. Additionally, the output field in the assembly instruction is left blank, as store instructions do not have any outputs.

 The assembler files, particularly the opcode table, were modified to support the new inline assembly instructions. These table modifications follow the same structure as the entries for standard store instructions, with the primary difference being the match field. This adjustment ensures that the match field sets the appropriate minor opcode bits, which differ from the standard store instructions, in the machine code.

After ensuring the machine code for the new instructions was correctly set, changes were introduced in the HDL to decode and implement the functionality of the new instructions. In the decode stage, four hot bits, covering all fast store granularities, were added to the instruction vector. Once decoded, the new instructions follow a similar path as ordinary store operations in terms of setting critical flags, such as the store/read flag and memory size indicator. 
To propagate fast store status from execute stage to cache stage, a fast store flag was added to a propagating package.

\begin{figure}
    \centering
    \includegraphics[width=0.655\linewidth]{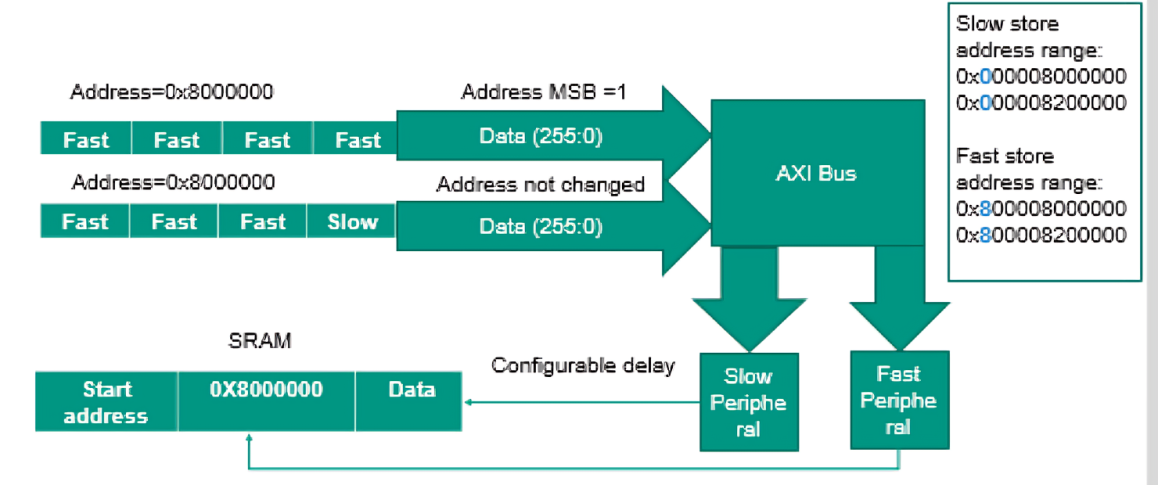}
    \caption{Cache behavior for slow and fast writes. If all words written in a cache line can use fast write, it is written back to memory using fast write mode. Otherwise, it uses slow write mode. Slow mode and fast mode each has a different address range.}
    \label{fig :CacheBehavior}
\end{figure}

\subsubsection{Cache Stage Modification in VHDL}
To be able to use the new memory instruction effectively, modification of Cache is needed.
The cache uses a write-back policy with each line consisting of four double words and follows an incremental burst strategy during write-back or flushing. It includes standard flags like dirty, valid, and shared. Four additional fast store flags were added to these flags, where each one indicates fast write status for each double word in the cache line.
If all fast store flags are set, the entire cache line is forwarded to the fast store peripheral by setting the address MSB to 1. Otherwise, the cache line is sent to the slow store peripheral by leaving the address unchanged.

An illustration of the typical cache behavior for fast and slow writes is provided in Figure~\ref{fig :CacheBehavior}.

\subsection{RISC-V Extension for Bit Manipulation} 
\label{sub_sec:Bit Manipulation}

\purple{Many applications include data where certain bits are more critical and demand higher retention than others. To support such cases, we enable fine-grained control over non-volatile memory (NVM) behavior—particularly bit-level manipulation for retention-aware and asymmetric-write scenarios—by extending our RISC-V SoC with a custom AXI-based, memory-mapped peripheral. Conventional memory interfaces, which apply uniform write modes (e.g., fast or slow) to entire words, cannot efficiently handle such asymmetric requirements.}

\purple{Our peripheral addresses this limitation by supporting selective bit-level control. It collates writes to consecutive memory addresses and internally transposes the data, aligning bits at the same position across multiple words. This transformation allows each bit column to be written using a distinct write mode, enabling retention-aware or performance-optimized access. On read, the peripheral applies the inverse transposition to maintain a consistent memory view for the processor.}

\purple{The peripheral is fully reconfigurable and supports both the normal and transposed memory views, allowing flexible software exploration of NVM behaviors such as selective bit reads, retention-aware transformations, and dynamic reconfiguration. These capabilities are discussed in further detail below.}

\begin{figure}
\centering
\includegraphics[page = 1,width=0.97\textwidth, clip, trim=0.5cm 1.5cm 0.5cm 2.9cm]{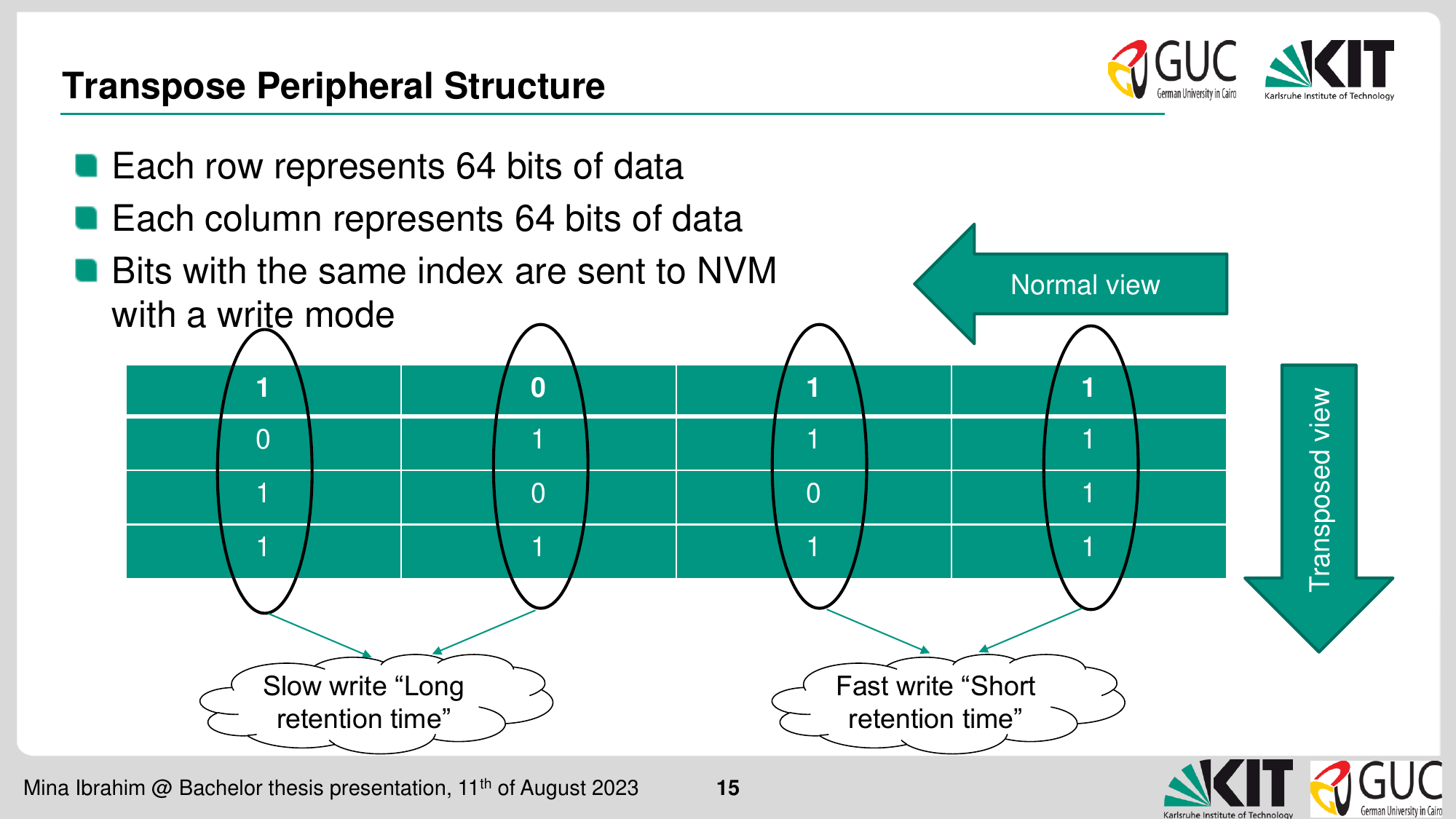}
\caption{Example illustrating peripheral structure and function.}
\label{fig:Example to understand peripheral structure and function}
\end{figure}

\purple{Internally, the peripheral behaves like a 64×64 matrix (see Figure~\ref{fig:Example to understand peripheral structure and function}) and supports two distinct data access layouts: a normal view (row-wise access) and a transposed view (column-wise access). The access mode can be configured independently for read and write paths using two one-bit configuration registers. A value of 0 selects the normal view, while 1 selects the transposed view. This allows any combination of access patterns—e.g., writing in normal view and reading in transposed view, or vice versa—enabling flexible retention-aware memory strategies.}

\purple{This configurability is particularly beneficial for emulating NVM characteristics where certain bit positions—such as the first column across all rows—require slow-write modes for longer retention, while others benefit from fast-write modes. By using the transposed view, the peripheral groups all bits of a specific column and applies uniform write strategies, regardless of the original row-wise access pattern.}

\purple{The peripheral operates as a matrix buffer with independent control logic for read and write operations across either view. While this design provides rich configurability and supports in-memory transformation experiments, it increases logic utilization, especially as the matrix size grows. We analyze this trade-off between hardware overhead and data-access flexibility in Section~\ref{sec:Bit Manipulation}.}

\purple{To ensure consistency between views, the peripheral is marked non-cacheable via the RISC-V platform's Physical Memory Attribute (PMA) configuration. This guarantees that all loads fetch data directly from the peripheral and all stores bypass the cache. Without this constraint, local cached data written in one view could be erroneously reused in another (processor), causing inconsistencies.}

\subsection{Experimental Setup}  
After designing the memory controller and extending RISC-V with new instructions, we built an experimental setup to measure access latency for various NVM technologies and verify the RISC-V extension.
\subsubsection{Memory Controller Testing}  
\begin{figure}
    \centering
    \includegraphics[width=0.65\linewidth]{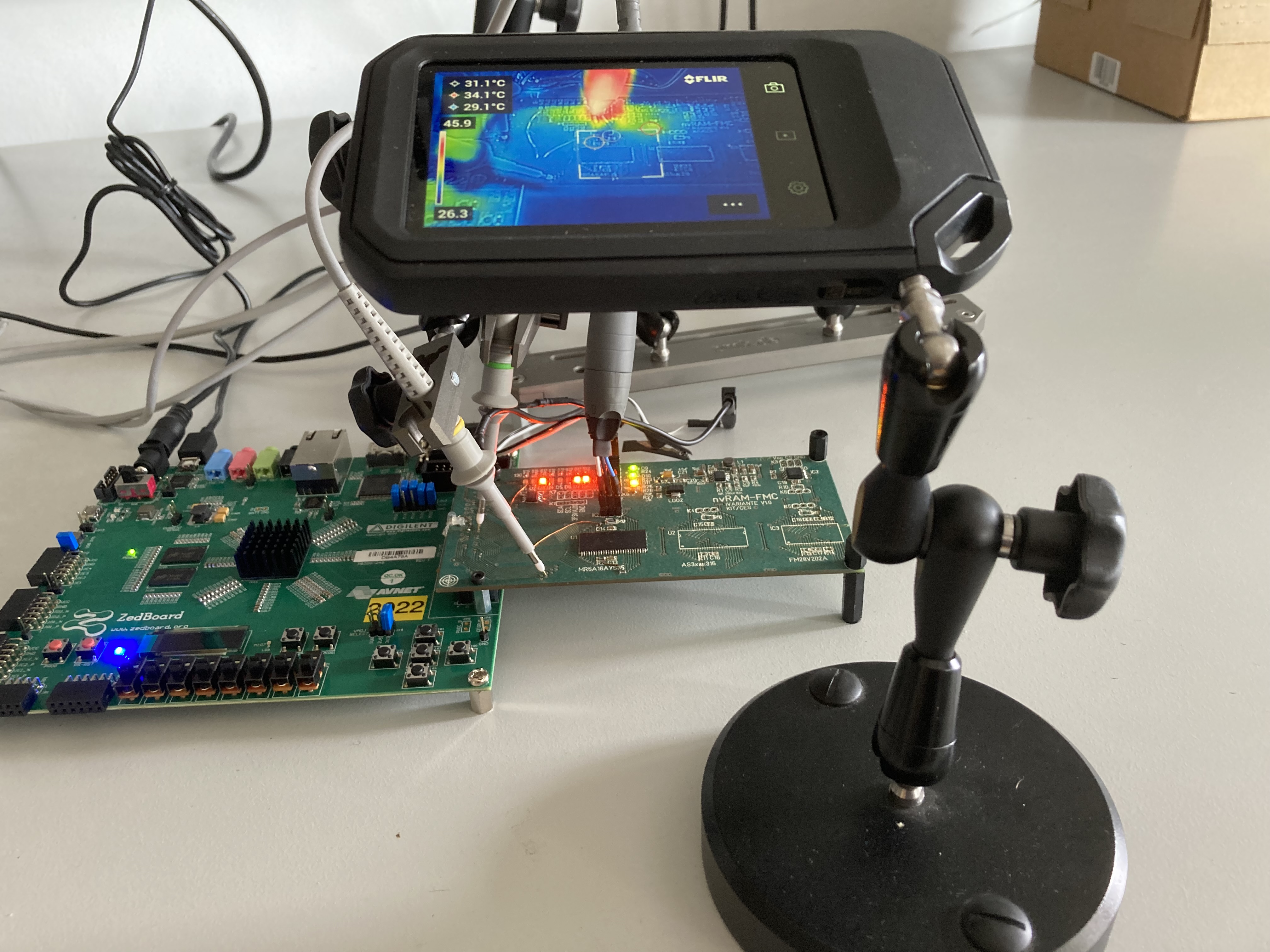}
    \caption{NVM performance measurement setup. The developped PCB is connected to a Zedboard via FMC. Oscilloscope probes are connected to collect the results and a thermal camera is used to ensure that no overheating occurs.}
    \label{fig:RealSetup}
\end{figure}

We implemented the system in VHDL and tested our memory controller, focusing on access latency for different NVMs, including ReRAM, FRAM, and MRAM. ReRAM used a serial SPI interface, while FRAM and MRAM required an 8-bit parallel interface. We used interfaces provided by the commercial NVM chips (Table~\ref{tab:purchased_rams}). Figure~\ref{fig:RealSetup} shows the general setup. ReRAM was mounted on an adapter board connected to a Zedboard via FMC pins. The processor executed C test programs implemented as console applications initiated by user input. Most tests ran automatically, generating target addresses and test data. A byte stream, composed of a control command, 3 address bytes, and 252 data bytes, was sent to the AXI\_to\_SPI IP block. Data passed through a 256-byte FIFO before transmission. Glue logic mapped SPI signals to the FMC pins, enabling data transfer from the FPGA to ReRAM, which also had a 256-byte FIFO for data. Power consumption was measured with an oscilloscope connected directly to the ReRAM.

\begin{table}
    \centering
  \caption{Selected and ordered nvRAMs}
    \begin{tabular}{|c|c|c|c|c|}
      \hline
    Manufacturer & Technology & Name & Size & Interface\\
        \hline
    Avalanche & STT MRAM & AS3004316 & 4 Mb & Parallel x16\\
        \hline
    Infineon & FRAM & FM28V202A-TG & 2 Mb & Parallel x16\\
        \hline
    Fujitsu & ReRAM & MB85AS8MT & 2 Mb & SPI (Serial)\\
        \hline
  \end{tabular}
  \label{tab:purchased_rams}
\end{table}

For speed measurements, after a write command was sent, the 'write in progress' bit of the status register was queried to confirm when the transfer from FIFO to NVM was complete. Performance varied by packet size and pattern, and a test program wrote data in different sizes and patterns to measure this. The CPU timed the write and read operations to calculate performance.

\subsubsection{Emulating the New RISC-V Instruction}  
\begin{figure}
    \centering
    \includegraphics[width=.5\linewidth]{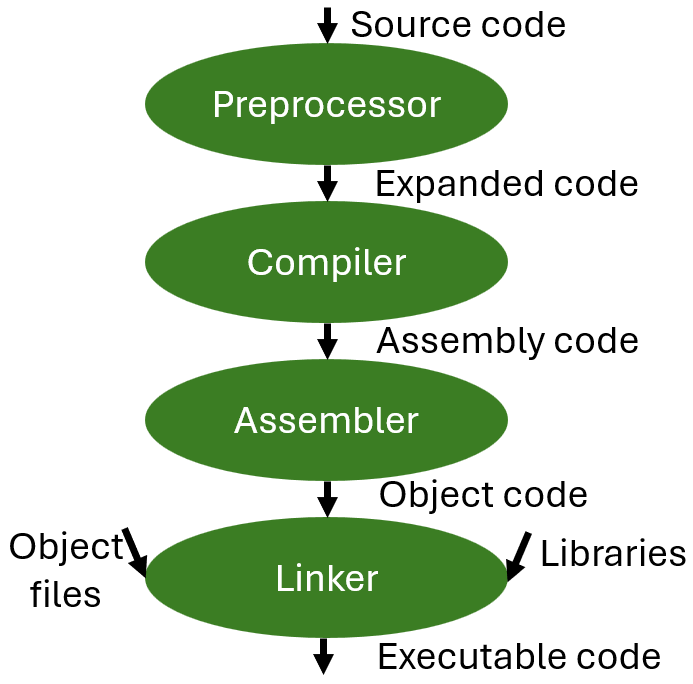}
    \caption{Compilation process}
    \label{fig:Compilation process}
\end{figure}

As the NVM chips did not provide access to different MLC write modes, the RISC-V write mode extension was emulated using RAM with added delays. To support our new instruction, we used the GCC toolchain compatible with RISC-V to compile, assemble, and link the program. \textit{Startup.s} and the provided RISC-V source codes were used in the process. The compiled executable was converted into two HEX files (upper and lower 32 bits) using \textit{elf2rawx}. These were uploaded to the RISC-V processor via HDL code, and the ROM was initialized with the program. Output was observed via UART serial communication. Figure~\ref{fig:Compilation process} illustrates the compilation process.
        
\section{Results} 
\label{sec:Results}
Our evaluation focuses on the hardware overhead and timing performance of the custom memory controller and new RISC-V instructions. We compare the custom memory controller with Xilinx's controller, assess the minimal impact of the new instructions, and review read/write latency for ReRAM, FRAM, and MRAM.
\subsection{Hardware Overhead}
We first analyze the hardware overhead of using our memory controller and adding the new instruction.
Adding the custom memory controller has saved considerable hardware cost as shown in Table~\ref{tab:HardwareUtilization2}. We have used Xilinx's auto-generated memory controller for reference. It introduced 339 additional flip flops and 429 lookup tables compared to 450 flip flops and 563 lookup tables of the built-in Xilinx controller, yielding a 30\% reduction in hardware cost (by developing a dedicated NVM memory specific controller). 
\begin{table}
    \centering
    \caption{Hardware Utilization of the Memory Controller}
    \begin{tabular}{|c|c|c|c|c|}
      \hline
       \textbf{Utilization of} & \textbf{LUTs} & \textbf{Registers} \\
       
        \hline
        Xilinx Built-in Memory Controller & 450 & 563 \\
        \hline
        Custom Memory Controller & 339 & 429 \\
        \hline
    \end{tabular}
    \label{tab:HardwareUtilization2}
\end{table}

Moreover, the hardware utilization results of adding the new instruction is shown in table~\ref{tab:Hardware Utilization1}, where the LUTs have increased by 0.03\% and registers are approximately the same.

\begin{table}
    \centering
    \caption{Hardware Utilization of the instruction set extension}
    \begin{tabular}{|c|c|c|c|c|}
      \hline
       \textbf{Utilization of RISC-V in \%} & \textbf{LUTs} & \textbf{Registers} 
       \\
        \hline
        Initially & 40.59 & 10.16 \\
        \hline
        After adding new instruction & 40.56 & 10.19   \\
        \hline

    \end{tabular}
    \label{tab:Hardware Utilization1}
\end{table}

\subsection{Timing Analysis}
\begin{figure}
    \centering
    \includegraphics[width=0.9\linewidth]{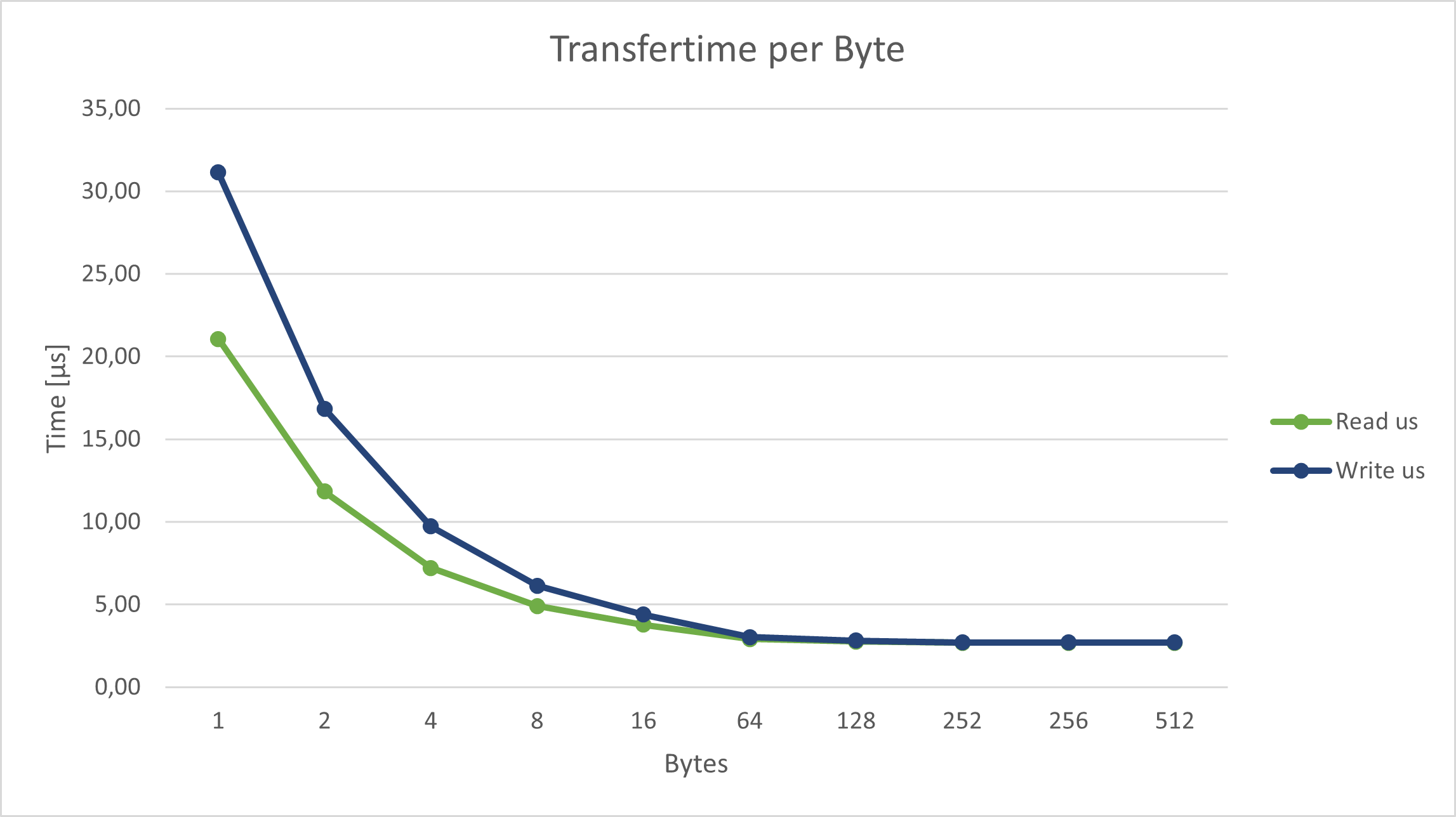}
    \caption{Transfer Time for Different Packet Sizes}
    \label{fig:transferTime}
\end{figure}

Next, we investigate the delay of using the different NVM commercial chips.
Regarding ReRAM speed performance, Figure~\ref{fig:transferTime} shows how packet size affects the write and read times. Time per byte is decreasing with the increasing number of bytes in a packet due to the huge overhead of the SPI commands with additional commands for activation and polling. After reaching the FIFO limit of 252 bytes, the packets need to split as no improvements are possible.

\begin{table}
    \centering
    \caption{Access Time Per Byte}
    \begin{tabular}{|c|c|c|c|c|}
      \hline
       \textbf{Type Of Memory} & \textbf{FRAM} & \textbf{MRAM} & \textbf{ReRAM}  \\
       
        \hline
        Read Time & 61 ns & 41 ns & 1.6 us \\
        \hline
        Write Time & 86 ns & 66 ns & 15 us \\
        \hline
    \end{tabular}
    \label{tab:Access Time Per Byte}
\end{table}

\begin{figure}
    \centering
    \includegraphics[width=0.8\linewidth]{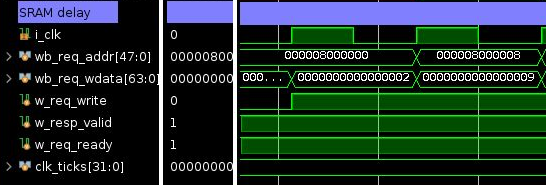}
    \caption{SRAM signals with no delay}
    \label{fig:SRAM signal with no delay}
\end{figure}

\begin{figure}
    \centering
    \includegraphics[trim={0cm 0cm 4cm 0cm},clip,scale = 0.7]{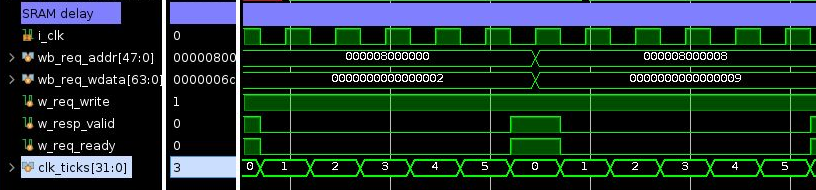}
    \caption{SRAM signals with delay equal to 5 clock cycles which can be observed from the clk\_ticks signal which means that the write response valid and write request ready are delayed by 5 clock cycles to delay the next write instruction.}
    \label{fig:SRAM signals with delay equal to 5 clock cycles}
\end{figure}

For brevity we do not show the curves of the other NVM chips but rather the ReRAM performance results are compared to FRAM and MRAM, as shown in Table~\ref{tab:Access Time Per Byte}. The table presents the access time per byte for 64-byte transactions.

As discussed in Experimental setup, a configurable delay is implemented in the SRAM to emulate the MLC behavior. Figure~\ref{fig:SRAM signal with no delay} and Figure~\ref{fig:SRAM signals with delay equal to 5 clock cycles} demonstrate this behavior. Further, we excecuted a streaming workload with both slow and fast store variants. Cache hits were observed for accesses within the cache line width, while the remaining accesses were serviced via main memory. The fast-store workload outperformed the slow-store workload by 7.7\%. As mentioned earlier, we emulated the slow write with a 5-cycle delay. This cycle difference is configurable, and users can adjust the delay to suit specific memory types.

\begin{figure}
    \centering
    \includegraphics[scale = 0.8]{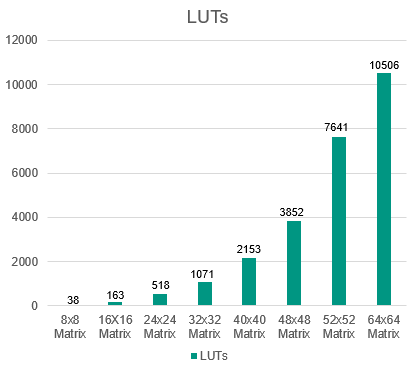}
    \caption{LUTs pattern for AXI Transpose Peripheral}
    \label{fig:LUTs pattern for AXI Transpose Peripheral}
\end{figure}

\begin{figure}
    \centering
    \includegraphics[scale = 0.27]{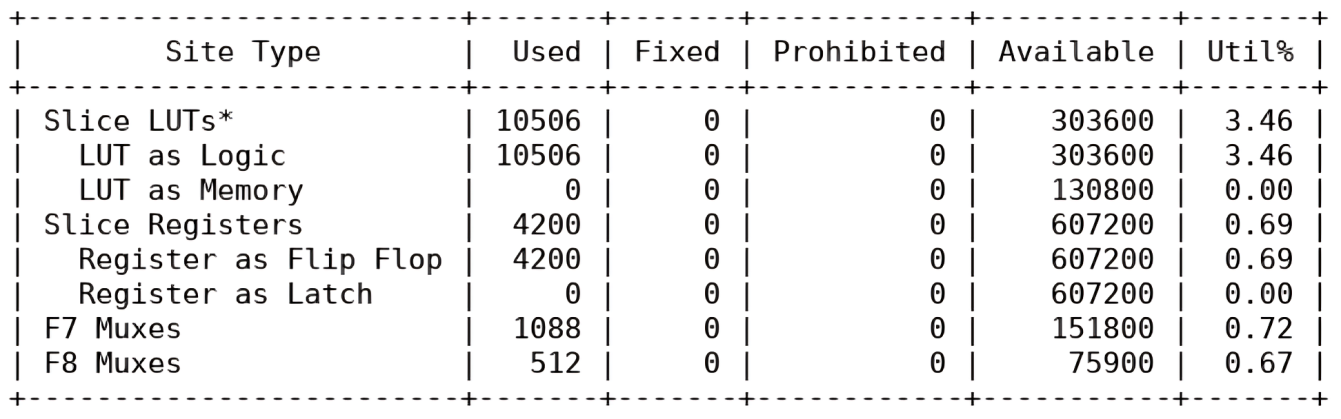}
    \caption{Transpose peripheral (64 x 64) synthesis report}
    \label{fig:Transpose peripheral synthesis report}
\end{figure}

\subsection{Bit Manipulation}
\label{sec:Bit Manipulation}

\purple{The transpose peripheral enables bit-level control over write operations, offering a fine-grained interface to optimize memory access patterns. As matrix dimensions increase, hardware utilization rises accordingly due to the growth in control logic and data routing complexity. Figure~\ref{fig:LUTs pattern for AXI Transpose Peripheral} shows a steep increase in LUT usage with matrix size, from only 38 LUTs for an 8×8 matrix to 10,506 LUTs for a 64×64 matrix.}

\purple{Despite this growth, the overall footprint remains modest. Even at 64×64, the design utilizes just 3.46\% of the available LUTs on the target FPGA, indicating ample headroom for integration into larger designs or SoCs. For mid-sized configurations such as 32×32, the LUT overhead is under 1\%, making it suitable for area-constrained applications. The complete synthesis summary is shown in Figure~\ref{fig:Transpose peripheral synthesis report}, confirming low utilization across other FPGA resources, such as registers and multiplexers.}

\purple{This efficient hardware profile, combined with the ability to specify write modes at bit-level granularity, provides a scalable and configurable solution for write-intensive applications.}

\section{Conclusion}
\label{sec:Conclusion}
This paper presented an innovative approach to improving MLC NVM performance through a custom NVM memory controller and RISC-V instruction set extension. Our controller, utilizing a finite state machine and AXI interface, enables efficient memory access with burst transfer capabilities, while minimizing hardware overhead. The proposed fast-store instruction reduces write latency by addressing the trade-off between write speed and retention time in MLC NVMs. \purple{Furthermore, we introduced a retention-aware write architecture that distinguishes between critical and non-critical data, applying slower, high-retention writes to the most significant bits (MSBs) and faster, low-retention writes to the least significant bits (LSBs), thereby optimizing both retention and performance.} FPGA-based experiments with multiple NVMs confirmed the design’s performance improvements in speed and resource usage. These contributions enhance MLC NVM efficiency in embedded systems and memory-intensive applications, laying the groundwork for future optimization of NVM architectures.

\section{Acknowledgments}
This work is supported by German Research Foundation (DFG): ARTS-NVM (part of SPP2377) and NVM-OMA (Project Number: 405422836) projects.

\bibliographystyle{ieeetr}  
\bibliography{sample}       

@article{kim2016improving,
  title={Improving write performance by controlling target resistance distributions in MLC PRAM},
  author={Kim, Youngsik and Yoo, Sungjoo and Lee, Sunggu},
  journal={ACM Transactions on Design Automation of Electronic Systems (TODAES)},
  volume={21},
  number={2},
  pages={1--27},
  year={2016},
  publisher={ACM New York, NY, USA}
}

@inproceedings{pan20143m,
  title={3M-PCM: Exploiting multiple write modes MLC phase change main memory in embedded systems},
  author={Pan, Chen and Xie, Mimi and Hu, Jingtong and Chen, Yiran and Yang, Chengmo},
  booktitle={Proceedings of the 2014 International Conference on Hardware/Software Codesign and System Synthesis},
  pages={1--10},
  year={2014}
}

@article{pan2017exploiting,
  title={Exploiting multiple write modes of nonvolatile main memory in embedded systems},
  author={Pan, Chen and Xie, Mimi and Yang, Chengmo and Chen, Yiran and Hu, Jingtong},
  journal={ACM Transactions on Embedded Computing Systems (TECS)},
  volume={16},
  number={4},
  pages={1--26},
  year={2017},
  publisher={ACM New York, NY, USA}
}

@article{khwa2016retention,
  title={A Retention-Aware Multilevel Cell Phase Change Memory Program Evaluation Metric},
  author={Khwa, Win-San and Chang, Meng-Fan and Wu, Jau-Yi and Lee, Ming-Hsiu and Su, Tzu-Hsiang and Wang, Tien-Yen and Li, Hsiang-Pang and BrightSky, Matthew and Kim, SangBum and Lung, Hsiang-Lan and others},
  journal={IEEE Electron Device Letters},
  volume={37},
  number={11},
  pages={1422--1425},
  year={2016},
  publisher={IEEE}
}

@inproceedings{ibrahim2024fpga,
  title={An FPGA-Based RISC-V Instruction Set Extension and Memory Controller for Multi-Level Cell NVM},
  author={Ibrahim, Mina and Shokry, Martel and Siddhu, Lokesh and Bauer, Lars and Nassar, Hassan and Henkel, J{\"o}rg},
  booktitle={2024 International Conference on Microelectronics (ICM)},
  pages={1--6},
  year={2024},
  organization={IEEE}
}

@article{mittal2016reliability,
  title={Reliability tradeoffs in design of volatile and nonvolatile caches},
  author={Mittal, Sparsh and Vetter, Jeffrey S},
  journal={Journal of Circuits, Systems and Computers},
  volume={25},
  number={11},
  pages={1650139},
  year={2016},
  publisher={World Scientific}
}

@article{priya2020enhancing,
  title={Enhancing the Lifetime of a Phase Change Memory with Bit-Flip Reversal},
  author={Priya, Bhukya Krishna and Ramasubramanian, N},
  journal={Journal of Circuits, Systems and Computers},
  volume={29},
  number={14},
  pages={2050219},
  year={2020},
  publisher={World Scientific}
}

@article{benhadjyoussef2015enhancing,
  title={Enhancing a 32-bit processor core with efficient cryptographic instructions},
  author={Benhadjyoussef, Noura and Elhadjyoussef, Wajih and Machhout, Mohsen and Tourki, Rached and Torki, Kholdoun},
  journal={Journal of Circuits, Systems and Computers},
  volume={24},
  number={10},
  pages={1550158},
  year={2015},
  publisher={World Scientific}
}

@article{zhang2019quick,
  author={Zhang, Mingzhe and Zhang, Lunkai and Jiang, Lei and Chong, Frederic T. and Liu, Zhiyong},
  journal={IEEE TC}, 
  title={Quick-and-Dirty: An Architecture for High-Performance Temporary Short Writes in MLC PCM}, 
  year={2019}}

@ARTICLE{siddhu2023swift,
  author={Siddhu, Lokesh and Nassar, Hassan and Bauer, Lars and Hakert, Christian and Hölscher, Nils and Chen, Jian-Jia and Henkel, Joerg},
  journal={IEEE ESL}, 
  title={Swift-CNN: Leveraging PCM Memory’s Fast Write Mode to Accelerate CNNs}, 
  year={2023}}

@article{anv-puf,
author = {Nassar, Hassan and Bauer, Lars and Henkel, J\"{o}rg},
title = {{ANV-PUF: Machine-Learning-Resilient NVM-Based Arbiter PUF}},
year = {2023},
journal = {ACM TECS}
}

@ARTICLE{nvmWear23,
  author={Hölscher, Nils and Hakert, Christian and Nassar, Hassan and Chen, Kuan-Hsun and Bauer, Lars and Chen, Jian-Jia and Henkel, Jörg},
  journal={IEEE TCAD}, 
  title={{Memory Carousel: LLVM-Based Bitwise Wear-Leveling for Non-Volatile Main Memory}}, 
  year={2023}
}

@article{pcm_review_2010,
  author  = "Burr, Geoffrey W. and Breitwisch, Matthew J. and Franceschini, Michele and Garetto, Davide and Gopalakrishnan, Kailash and Jackson, Bryan and Kurdi, Bülent and Lam, Chung and Lastras, Luis A. and Padilla, Alvaro and Rajendran, Bipin and Raoux, Simone and Shenoy, Rohit S.",
  title   = "Phase change memory technology",
  journal = "Journal of Vacuum Science \& Technology B",
  volume  = "28",
  number  = "2",
  pages   = "223--262",
  year    = "2010"
}

@INPROCEEDINGS{nassar_icm_23,
  author={Nassar, Hassan and Youssef, Rafik and Bauer, Lars and Henkel, Jörg},
  booktitle={ICM}, 
  title={Supporting Dynamic Control-Flow Execution for Runtime Reconfigurable Processors}, 
  year={2023}
}

@article{sttramMLC1,
author = {Pan, Chen and Xie, Mimi and Yang, Chengmo and Chen, Yiran and Hu, Jingtong},
title = {Exploiting Multiple Write Modes of Nonvolatile Main Memory in Embedded Systems},
year = {2017},
journal = {ACM TECS}
}

@inbook{mram_fram,
  author    = {Scott, J. F.},
  title     = {A Comparison of Magnetic Random Access Memories (MRAMs) and Ferroelectric Random Access Memories (FRAMs)},
  booktitle = {Ferro- and Antiferroelectricity: Order/Disorder versus Displacive},
  year      = {2007},
  publisher = {Springer},
  address   = {Berlin, Heidelberg},
  pages     = {199-207}
}

@Article{rram_phys_model,
author ="Chen, Ying-Chen and Chang, Yao-Feng and Wu, Xiaohan and Zhou, Fei and Guo, Meiqi and Lin, Chih-Yang and Hsieh, Cheng-Chih and Fowler, Burt and Chang, Ting-Chang and Lee, Jack C.",
title  ="Dynamic conductance characteristics in HfOx-based resistive random access memory",
journal  ="RSC Adv.",
year  ="2017",
publisher  ="The Royal Society of Chemistry"}

@article{stt_mram_endurance_study,
  author  = {Kan, Jimmy J. and Park, Chando and Ching, Chi and Ahn, Jaesoo and Xie, Yuan and Pakala, Mahendra and Kang, Seung H.},
  journal = {IEEE Transactions on Electron Devices},
  title   = {A Study on Practically Unlimited Endurance of STT-MRAM},
  year    = {2017}
}

@inproceedings{stt_mram_endurance_tsmc,
  author    = {Yu, Hung-Chang and Lin, Kai-Chun and Lin, Ku-Feng and Huang, Chin-Yi and Chih, Yu-Der and Ong, Tong-Chern and Chang, Jonathan and Natarajan, Sreedhar and Tran, Luan C.},
  booktitle = {International Solid-State Circuits Conference Digest of Technical Papers},
  title     = {Cycling endurance optimization scheme for 1Mb STT-MRAM in 40nm technology},
  year      = {2013}
}

@inproceedings{qureshi2009enhancing,
author = {Qureshi, Moinuddin K. and Karidis, John and Franceschini, Michele and Srinivasan, Vijayalakshmi and Lastras, Luis and Abali, Bulent},
title = {Enhancing lifetime and security of PCM-based main memory with start-gap wear leveling},
year = {2009},
booktitle = {MICRO}
}

@INPROCEEDINGS{riscv1,
  author={Nguyen-Hoang, Duc-Thinh and Ma, Khai-Minh and Le, Duy-Linh and Thai, Hong-Hai and Cao, Tran-Bao-Thuong and Le, Duc-Hung},
  booktitle={IEEE ICCE}, 
  title={Implementation of a 32-Bit RISC-V Processor with Cryptography Accelerators on FPGA and ASIC}, 
  year={2022}}

@INPROCEEDINGS{riscv2,
  author={P, Arul and N, Abirami and S, Sayeekumar and P, Vanmathi and V, Annapoorani},
  booktitle={ICAECT}, 
  title={Implementation of RISC-V Instruction Set Architecture for edge IoT computing platform}, 
  year={2024}}

@inproceedings{li2013compiler,
  title={Compiler directed write-mode selection for high performance low power volatile {PCM}},
  author={Li, Qingan and Jiang, Lei and Zhang, Youtao and He, Yanxiang and Xue, Chun Jason},
  booktitle={LCTES},
  year={2013},
  pages__={101--110},
}

@article{qiu2015write,
  title={Write mode aware loop tiling for high performance low power volatile PCM in embedded systems},
  author={Qiu, Keni and Li, Qingan and Hu, Jingtong and Zhang, Weigong and Xue, Chun Jason},
  journal={TC},
  year={2015},
  publisher={IEEE}
}

@INPROCEEDINGS{riscv3,
  author={Kanjarbhat, Prem and Chandaragi, Pooja and Wali, Uday V},
  booktitle={AMATHE}, 
  title={Implementation of PWM Using RISC-V Processor}, 
  year={2024}}

@misc{Waterman_Asanovic_2017, title={The {RISC-V} instruction set manual}, url={https://riscv.org/wp-content/uploads/2017/05/riscv-spec-v2.2.pdf}, journal={The RISC-V Instruction Set Manual}, author={Waterman, Andrew and Asanovic, Krste}, year={2017}, month={May}}

\end{document}